\documentclass[12pt]{article}
 \usepackage{epsfig}
 \def\be{\begin{equation}}
 \def\ee{\end{equation}}
 \def\bea{\begin{eqnarray}}
 \def\eea{\end{eqnarray}}
 \usepackage{graphicx}

 \catcode`\@=11
 \def\lsim{\mathrel{\mathpalette\@versim<}}
 \def\gsim{\mathrel{\mathpalette\@versim>}}
 \def\@versim#1#2{\vcenter{\offinterlineskip
 \ialign{$\m@th#1\hfil##\hfil$\crcr#2\crcr\sim\crcr } }}
 \catcode`\@=12

 \catcode`@=12
\begin{document}
 \thispagestyle{empty}
 \begin{flushright}
 UCRHEP-T585\\
 Oct 2017\
 \end{flushright}
 \vspace{0.6in}
 \begin{center}
 {\LARGE \bf Dark Revelations of the $[SU(3)]^3$ and 
 $[SU(3)]^4$\\
Gauge Extensions of the Standard Model\\}
 \vspace{1.2in}
 {\bf Corey Kownacki, Ernest Ma, Nicholas Pollard, Oleg Popov, and 
Mohammadreza Zakeri\\}
 \vspace{0.2in}
 {\sl Physics and Astronomy Department,\\ 
 University of California, Riverside, California 92521, USA\\}
 \end{center}
 \vspace{1.2in}

\begin{abstract}\
Two theoretically well-motivated gauge extensions of the standard 
model are $SU(3)_C \times SU(3)_L \times SU(3)_R$ and 
$SU(3)_q \times SU(3)_L \times SU(3)_l \times SU(3)_R$, where $SU(3)_q$ is 
the same as $SU(3)_C$ and $SU(3)_l$ is its color leptonic counterpart. 
Each has three variations, according to how $SU(3)_R$ is broken.  It is 
shown here for the first time that a built-in dark $U(1)_D$ gauge symmetry 
exists in all six versions, and may be broken to discrete $Z_2$ dark parity.  
The available dark matter candidates in each case include fermions, scalars, 
as well as {\it vector gauge bosons}.  This work points to the unity 
of matter with dark matter, the origin of which is not {\it ad hoc}.

\end{abstract}

 \newpage
 \baselineskip 24pt
\noindent \underline{\it Introduction}~:~
To extend the $SU(3)_C \times SU(2)_L \times U(1)_Y$ gauge symmetry of the 
standard model (SM) of quarks and leptons, there are many possibilities. 
We focus in this paper on two such theoretically well-motivated ideas. 
The first~\cite{dgg84,bhp86} is $SU(3)_C \times SU(3)_L \times SU(3)_R$, and 
the second~\cite{jv92,bmw04,kmppz17} is 
$SU(3)_q \times SU(3)_L \times SU(3)_l \times SU(3)_R$, where $SU(3)_q$ is the 
same as $SU(3)_C$ and $SU(3)_l$ is its color leptonic counterpart.  It has 
been known for a long time that $[SU(3)]^3$ has three distinct variations, 
according to how $SU(3)_R$ is broken to $SU(2)_R$.  
\begin{itemize}
\item{(A) $(u,d)_R$ is 
a doublet, which corresponds to the conventional left-right model.}
\item{(B) $(u,h)_R$ is a doublet~\cite{m87,bhm87,klm09,klm10,bmw14,kmppz17-1}, 
where $h$ is an exotic quark with the same charge as $d$, which corresponds 
to the alternative left-right model.} 
\item{(C) $(h,d)_R$ is a doublet~\cite{lr86,dm11,bdmw12,fmz15}, which implies 
that the vector gauge bosons of this $SU(2)_R$ are all neutral.}
\end{itemize}
Note that in the early days of flavor $SU(3)$ for the $u,d,s$ quarks, 
these $SU(2)$ subgroups are called $T,V,U$ spins.  The same three versions 
are obviously also possible for $[SU(3]^4$.

Whereas these structures have been known for a long time, an important 
property of these models has been overlooked, i.e. the existence of a 
built-in dark $U(1)_D$ gauge symmetry already present in $[SU(3)]^3$ and 
$[SU(3)]^4$ under which the SM particles are distinguished from those of 
the dark sector.  We will identify this symmetry in all six cases and 
discuss how it may fit into a viable extension of the SM.

\noindent \underline{\it Dark Symmetries in $[SU(3)]^3$}~:~
The fermion assignments under $SU(3)_C \times SU(3)_L \times SU(3)_R$ 
are
\begin{eqnarray}
q \sim (3,3^*,1) \sim \pmatrix{d & u & h \cr d & u & h \cr d & u & h},
\end{eqnarray}
where the $I_{3L}$ values from left to right are $(-1/2,1/2,0)$ and the $Y_L$ 
values from left to right are $(-1/3,-1/3,2/3)$;
\begin{eqnarray}
\lambda \sim (1,3,3^*) \sim \pmatrix{N & E^c & \nu \cr E & N^c & e \cr 
\nu^c & e^c & S},
\end{eqnarray}
where the $I_{3L}$ values from top to bottom are now $(1/2,-1/2,0)$ and 
the $Y_L$ values from top to bottom are $(1/3,1/3,-2/3)$,  the  
$I_{3R}$ values from left to right are $(-1/2,1/2,0)$ and the $Y_R$ 
values from left to right are $(-1/3,-1/3,2/3)$;
\begin{eqnarray}
q^c \sim (3^*,1,3) \sim \pmatrix{d^c & d^c & d^c \cr u^c & u^c & u^c \cr h^c & 
h^c & h^c},
\end{eqnarray}
where the $I_{3R}$ values from top to bottom are $(1/2,-1/2,0)$ and the $Y_R$ 
values from top to bottom are $(1/3,1/3,-2/3)$.  The electric charge 
operator is given by
\begin{equation}
Q = I_{3L} - {Y_L \over 2} + I_{3R} - {Y_R \over 2}.
\end{equation}

Since $(d^c,u^c)$ and $(e^c,\nu^c)$ are $SU(2)_R$ doublets, this reduces 
to the conventional left-right model.  Consider now
\begin{equation}
D_A = 3(Y_L - Y_R).
\end{equation}
The $[Q,D_A]$ assignments of $q$, $\lambda$, and $q^c$ are then given by
\begin{eqnarray}
&& Q_q = \pmatrix{-1/3 & 2/3 & -1/3 \cr -1/3 & 2/3 & -1/3 \cr -1/3 & 2/3 & 
-1/3},  ~~~ D_q = \pmatrix{-1 & -1 & 2 \cr -1 & -1 & 2 \cr -1 & -1 & 2}, \\  
&& Q_\lambda = \pmatrix{0 & 1 & 0 \cr -1 & 0 & -1 \cr 0 & 1 & 0}, ~~~ 
D_\lambda = \pmatrix{2 & 2 & -1 \cr 2 & 2 & -1 \cr -1 & -1 & -4}, \\ 
&& Q_{q^c} = \pmatrix{1/3 & 1/3 & 1/3 \cr -2/3 & -2/3 & -2/3 \cr 1/3 & 1/3 
& 1/3}, ~~~ D_{q^c} = \pmatrix{-1 & -1 & -1 \cr -1 & -1 & -1 \cr 2 & 2 & 2}.
\end{eqnarray}
This shows that $u,u^c,d,d^c,\nu,\nu^c,e,e^c$ have $D_A=-1$ (odd), whereas 
$h,h^c,N,N^c,E,E^c,S$ have even $D_A$ charges, i.e. $2$ 
and $-4$.  Let us define a parity~\cite{m15} using the particle's spin $j$: 
\begin{equation}
R_A = (-1)^{D_A + 2j}.
\end{equation}
Since $j=1/2$, $R_A$ is 
even for $u,u^c,d,d^c,\nu,\nu^c,e,e^c$ and odd for $h,h^c,N,N^c,E,E^c,S$, 
thereby allowing the latter to be considered as belonging to the dark sector, 
as long as $U(1)_D$ is broken only by two units, in analogy to the 
breaking of $B-L$ in models of neutrino mass, where lepton parity $(-1)^L$ 
remains conserved.

To break $[SU(3)]^3$, a scalar bitriplet $\phi \sim (1,3,3^*)$ is used. 
It transforms exactly as $\lambda$ and has the same $[Q,D]$ assignments. 
Now $\langle \phi_{33} \rangle$ breaks $SU(3)_L \times SU(3)_R$ to 
$SU(2)_L \times SU(2)_R \times U(1)_{Y_L+Y_R}$.  The $U(1)_D$ symmetry 
is broken by 4 units at the same time.  This gives masses to the exotic 
fermions $h,N,E$.  Two other neutral scalars $\phi_{11},\phi_{22}$ have 
$D_A=2$.  Their vacuum expectation values would break 
$SU(2)_L \times SU(2)_R$ to $U(1)_{I_{3L}+I_{3R}}$, and 
$U(1)_D$ by 2 units, allowing mass terms for $uu^c$, $dd^c$, $ee^c$, 
$\nu \nu^c$, $NS$, and $N^c S$.  At this point, it looks like a dark residual 
$Z_2$ symmetry is still possible.  However this is not a viable 
scenario, because the $SU(2)_L$ and $SU(2)_R$ breaking are now at 
the same scale, contrary to what is observed.  Furthermore, both 
$I_{3L}+I_{3R}$ and $Y_L+Y_R$ are still unbroken.  Whereas $Q$ is a linear 
combination of the two, there remains another unbroken $U(1)$ gauge symmetry. 
To solve these problems, the usual procedure is to allow $\phi_{31}$ and 
$\phi_{13}$ to acquire nonzero vacuum expectation values as well, thus 
breaking $SU(2)_R$ and $SU(2)_L$ separately. However, 
since they have $D_A = -1$ (odd $R_A$), the dark symmetry is lost.

To save the dark symmetry, we insert another bitriplet $\eta \sim (1,3,3^*)$ 
with an extra $Z_2$ symmetry under which it is odd and all other fields 
are even. This extra symmetry prevents $\eta$ from coupling to the quarks and 
leptons, so that the absolute $R_A$ values of the $\eta$ components are not 
fixed by them as in $\phi$.  However their relative $R_A$ values are still 
fixed by the gauge bosons.  Using Eqs.~(5) and (9), we see that of the eight 
$SU(3)_L$ and eight $SU(3)_R$ gauge bosons, the four gauge bosons 
which take $u$ and $d$ to $h$, and the corresponding ones which take $u^c$ and 
$d^c$ to $h^c$ are odd under $R_A$, and the others are even.  We can now 
choose $\langle \eta_{31} \rangle \neq 0$ and $\langle \eta_{13} \rangle \neq 0$ 
to break $SU(3)_L \times SU(3)_R$ to just $U(1)_Q$ and preserve $R_A$, 
because $\eta_{31},\eta_{32},\eta_{13},\eta_{23}$ may be defined to be even 
and the other components odd without breaking $R_A$.

Of the 27 fermion fields for each family, 16 are in the visible sector 
($R_A$ even), i.e. $u,u^c,d,d^c,\nu,\nu^c,e,e^c$, and 11 are in the dark 
sector ($R_A$ odd), i.e. $h,h^c,N,N^c,E,E^c,S$.  Of the 24 gauge bosons, 
16 are visible, i.e. the 8 gluons, $W_L^\pm$, $W_R^\pm$, the photon, $Z$, 
and two other heavier neutral ones, a linear combination of which 
couples to the dark charge $D_A$, and 8 are dark, i.e. those with 
odd $R_A$.  The scalars are also divided into sectors with 
even and odd $R_A$.  This is thus a model with possible fermion, scalar, 
and {\it vector} dark-matter candidates.  Their existence is not an 
{\it ad hoc} 
invention, but a possible outcome of the postulated theoretical framework 
beyond the standard model.

Consider next the alternative left-right model, i.e. variation (B), 
where $d^c$ is switched 
with $h^c$ and $(\nu,e,S)$ are switiched with $(N,E,\nu^c)$, i.e.
\begin{equation}
q^c \sim \pmatrix{h^c & h^c & h^c \cr u^c & u^c & u^c \cr d^c & d^c & d^c}, ~~~ 
\lambda \sim \pmatrix{\nu & E^c & N \cr e & N^c & E \cr S & e^c & \nu^c}.
\end{equation}
The electric charge is given as before by Eq.~(4), but the dark charge is now
\begin{equation}
D_B = 3(Y_L + I_{3R} + {Y_R \over 2}).
\end{equation}
Hence $D_q$ remains the same as in Eq.~(6), but $D_\lambda$ and $D_{q^c}$ are 
now given by
\begin{equation}
D_\lambda = \pmatrix{-1 & 2 & 2 \cr -1 & 2 & 2 \cr -4 & -1 & -1}, ~~~ 
D_{q^c} = \pmatrix{2 & 2 & 2 \cr -1 & -1 & -1 \cr -1 & -1 & -1}.
\end{equation}
Again using $R_{B} = (-1)^{D_B + 2j}$, we find it to be even for 
$u,u^c,d,d^c,\nu,\nu^c,e,e^c$ and odd for $h,h^c,N,N^c,E,E^c,S$. 
Choosing $\phi_{13},\phi_{22},\phi_{31}$ to have nonzero vacuum 
expectation values, the symmetry breaking pattern is as in (A), only 
that the $SU(2)$ subgroup of $SU(3)_R$ is now different.  It suffers 
from the same problems as in (A), which may be solved again by 
adding $\eta$, with $\langle \eta_{33} \rangle \neq 0$ and 
$\langle \eta_{11} \rangle \neq 0$.

In the third variation (C), $u^c$ is switched with $h^c$, and $(\nu,e,S)$ are 
switched with $(E^c, N^c,e^c)$, i.e.
\begin{equation}
q^c \sim \pmatrix{d^c & d^c & d^c \cr h^c & h^c & h^c \cr u^c & u^c & u^c}, ~~~ 
\lambda \sim \pmatrix{N & \nu & E^c \cr E & e & N^c \cr \nu^c & S & e^c}.
\end{equation}
The electric charge and dark charge are now given by
\begin{equation}
Q = I_{3L} - {Y_L \over 2} + Y_R, ~~~ D_C = 3(Y_L - I_{3R} + {Y_R \over 2}).
\end{equation}
Hence
\begin{eqnarray}
&& Q_\lambda = \pmatrix{0 & 0 & 1 \cr -1 & -1 & 0 \cr 0 & 0 & 1}, ~~~
D_\lambda = \pmatrix{2 & -1 & 2 \cr 2 & -1 & 2 \cr -1 & -4 & -1}, \\ 
&& Q_{q^c} = \pmatrix{1/3 & 1/3 & 1/3 \cr 1/3 & 1/3 & 1/3 \cr -2/3 & -2/3 & 
-2/3} ~~~ D_{q^c} = \pmatrix{-1 & -1 & -1 \cr 2 & 2 & 2 \cr -1 & -1 & -1}.
\end{eqnarray}
Again using $R_{C} = (-1)^{D_C + 2j}$, we find it to be even for 
$u,u^c,d,d^c,\nu,\nu^c,e,e^c$ and odd for $h,h^c,N,N^c,E,E^c,S$. 
Choosing $\phi_{11},\phi_{23},\phi_{32}$ to have nonzero vacuum 
expectation values, the pattern of symmetry breaking is the same 
as in (A) and (B), but the $SU(2)_R$ subgroup is different from either. 
It suffers from the same problems as the two previous cases, and they are 
again solved by adding $\eta$, with $\langle \eta_{31} \rangle \neq 0$ and 
$\langle \eta_{12} \rangle \neq 0$.  However, in contrast to the variations 
(A) and (B), the $\phi_{33}$ and $\eta_{33}$ entries are not neutral, so it 
is not possible to preserve $SU(2)_L \times SU(2)_R$ as a low-energy 
subgroup.

\noindent \underline{\it Gauge Boson Masses in (B)}~:~
Consider the breaking of $SU(3)_L \times SU(3)_R$ by a very large 
$\langle \eta_{33} \rangle = v_{33}$.  Of the 8 vector gauge bosons 
$W^L_i$ of $SU(3)_L$ and the 8 vector gauge bosons $W^R_i$ of $SU(3)_R$, 
9 become very heavy.  The remaining 7 are the 3 of $SU(2)_L$, the 3 of 
$SU(2)_R$, and the one linear combination $W^V_8 = (W^L_8+W^R_8)/\sqrt{2}$.  
We assume that they survive to just above the electroweak scale with 
equal couplings ($g$) for $SU(2)_L$ and $SU(2)_R$ and a different one ($g'$) 
for $Y_{L}+Y_{R}$.  Let $\langle \eta_{11} \rangle = v_{11}$, 
$\langle \phi_{22} \rangle = v_{22}$,  
$\langle \phi_{13} \rangle = v_{13}$,  
$\langle \phi_{31} \rangle = v_{31}$,  then
\begin{equation}
M^2(W^R_{1,2}) = {g^2 \over 2} [v_{11}^2 + v_{22}^2 + v_{31}^2],
\end{equation}
where $(W^R_1 \mp i W^R_2)/\sqrt{2} = W_R^\pm$ are the charged $SU(2)_R$ 
gauge bosons with odd $R_B$.  The other gauge bosons have even $R_B$ 
with 
\begin{equation}
M^2(W^L_{1,2}) = {g^2 \over 2} [v_{11}^2 + v_{22}^2 + v_{13}^2],
\end{equation}
and the massless photon given by
\begin{equation}
A = {e \over g} (W^L_3 + W^R_3) - {e \over g'} \sqrt{2 \over 3} W^V_8.
\end{equation}
This implies
\begin{equation}
{e^2 \over {g'}^2} = {3 \over 2} (1-2 \sin^2 \theta_W).
\end{equation}
If $g'=g$ (which is valid at the unification scale), then 
$\sin^2 \theta_W = 3/8$ as expected.  Now $v_{31}$ breaks $SU(2)_R$ 
without breaking $SU(2)_L$, so its value may be greater than the elctroweak 
scale.  Its associated gauge boson $Z'$ is given by
\begin{equation}
Z' = {\sqrt{2} g W^R_3 + \sqrt{3} g' W^V_8 \over \sqrt{2g^2 + 3 {g'}^2}} 
= {1 \over \cos \theta_W} [\sqrt{1-2\sin^2 \theta_W} W^R_3 + \sin \theta_W 
W^V_8].
\end{equation}
Hence the SM $Z$ boson is now
\begin{equation}
Z = \cos \theta_W W^L_3 - \tan \theta_W [\sin \theta_W W^R_3 - 
\sqrt{1-2\sin^2 \theta_W} W^V_8].
\end{equation}
The $(Z,Z')$ mass-squared matrix is given by
\begin{eqnarray}
M^2_{ZZ} &=& {g^2 \over 2 \cos^2 \theta_W} [v_{11}^2 + v_{22}^2 + v_{13}^2], 
\\  
M^2_{Z'Z'} &=& {g^2 \over 2} \left[ {\cos^2 \theta_W \over 1-2\sin^2 \theta_W} 
v_{31}^2 + {1-2 \sin^2 \theta_W \over \cos^2 \theta_W} (v_{11}^2 + v_{22}^2) 
+ 2 \tan^2 \theta_W v_{13}^2 \right], \\ 
M^2_{ZZ'} &=& {g^2 \tan^2 \theta_W \over 2 \sqrt{1-2\sin^2 \theta_W}} 
[\sin^2 \theta_W v_{13}^2 - (1-2\sin^2 \theta_W) (v_{11}^2 + v_{22}^2)].
\end{eqnarray}
To avoid $Z-Z'$ mixing so as not to upset precision electroweak measurements, 
$M^2_{ZZ'}$ may be chosen to be negligible in the above.

In this alternative left-right model, $(u,h)_R$ and $(S,e)_R$ are $SU(2)_R$ 
doublets with $h$ and $S$ odd under $R_B$.  The mass terms for $u$ and $\nu$ 
come from $v_{22}$, those for $d$ and $e$ from $v_{13}$, those for $h$, 
$E$ from $v_{31}$, and the $3 \times 3$ matrix spanning $(N,N^c,S)$ 
from all three.   As such, it contains the necessary ingredients for a 
consistent model of built-in dark matter.  In variation (C), it has 
already been noted that $SU(2)_L \times SU(2)_R$ cannot be maintained 
as a low-energy subgroup.  Hence the associated dark sector must be very 
heavy and does not lead to a realistic model.  In variation (A), whereas 
$SU(2)_L \times SU(2)_R$ may emerge as a low-energy subgroup, the 
dark sector consists of singlets under this symmetry and must also be 
very heavy.

\noindent \underline{\it Dark Symmetries in $[SU(3)]^4$}~:~
The notion of leptonic color~\cite{fl90,flv91} is based on quark-lepton 
interchange symmetry.  Postulating $SU(3)_l$ to go with $SU(3)_q$, leptons 
have three color components to begin with, but $SU(3)_l$ is broken to 
$SU(2)_l$ which remains exact, so that two of these leptonic color fields 
are confined in analogy to the three color quarks being confined.  The third 
unconfined component is the observed lepton of the SM.  The new particles of 
this model are not easily produced and observed at the Large Hadron 
Collider, but will have unique signatures in a future lepton collider, 
as recently discussed~\cite{kmppz17}.  Under 
$SU(3)_q \times SU(3)_L \times SU(3)_l \times SU(3)_R$, $q \sim (3,3^*,1,1)$ 
as in Eq.~(1) and $q^c \sim (3^*,1,1,3)$ as in Eqs.~(3), (10), and (13) for 
the three variations (A,B,C) in parallel to what has been discussed for 
$[SU(3]^3$.  As for the leptonic sector,
\begin{eqnarray}
l \sim (1,3,3^*,1) \sim \pmatrix{x_1 & x_2 & \nu \cr y_1 & y_2 & e \cr z_1 & 
z_2 & n}
\end{eqnarray}
is the same in all three variations, in analogy to $q$, whereas $l^c$ has 
three variations to match $q^c$, i.e.
\begin{equation}
l^c \sim (1,1,3,3^*) \sim \pmatrix{x_1^c & y_1^c & z_1^c \cr x_2^c & y_2^c 
& z_2^c \cr \nu^c & e^c & n^c}, ~~~ \pmatrix{z_1^c & y_1^c & x_1^c \cr 
z_2^c & y_2^c & x_2^c \cr n^c & e^c & \nu^c}, ~~~ 
\pmatrix{x_1^c & z_1^c & y_1^c \cr x_2^c & z_2^c & y_2^c \cr \nu^c & n^c 
& e^c}.
\end{equation}
The electric charge and dark charge in (A) are given by
\begin{equation}
Q = I_{3L} - {Y_L \over 2} + I_{3R} - {Y_R \over 2} - {Y_l \over 2}, ~~~ 
D_A = 3(Y_L - Y_R).
\end{equation}
Hence
\begin{equation}
Q_l = \pmatrix{1/2 & 1/2 & 0 \cr -1/2 & -1/2 & -1 \cr 1/2 & 1/2 & 0}, 
~~~ Q_{l^c} = \pmatrix{-1/2 & 1/2 & -1/2 \cr -1/2 & 1/2 & -1/2 \cr 
0 & 1 & 0},
\end{equation}
and $D_l = -D_{q^c}$ of Eq.~(8), $D_{l^c} = -D_{q}$ of Eq.~(6), i.e.   
$u,u^c,d,d^c,\nu,\nu^c,e,e^c,x,x^c,y,y^c$ have $D_A=1$ 
(odd), whereas $h,h^c,n,n^c,z,z^c$ have $D_A=-2$ (even).  Again 
let  $R_A = (-1)^{D_A + 2j}$, then the former group of fermions is even 
and the latter odd, i.e. belonging to the dark sector if $U(1)_D$ is broken 
only by two units.  

The breaking of $SU(3)_L \times SU(3)_R$ by a scalar bitriplet 
$\phi \sim (1,3,1,3^*)$, which couples also to the fermions, proceeds as 
before.  It has the same problems as discussed in the $[SU(3)]^3$ case. 
However, there are now two additional scalar bitriplets~\cite{bmw04} 
in $[SU(3)]^4$ with nonzero vacuum expectation values, i.e.
\begin{equation}
\phi^L \sim (1,3,3^*,1) \sim l, ~~~ \phi^R \sim (1,1,3,3^*) \sim l^c.
\end{equation}
They have thus the same would-be $[Q,D]$ assignments.  They are not 
responsible for fermion masses, but are required to break leptonic color 
$SU(3)_l$ to $SU(2)_l$.  Now $\phi^L_{33}$ has $D_A=2$ which may be used to 
break $SU(3)_l \times SU(2)_L$ to $SU(2)_l \times SU(2)_L \times U(1)_{Y_l+Y_L}$. 
To break $SU(2)_R$ as well without breaking $R_A$, we use the 
same trick as before by assigning $\phi^R$ an odd parity under $Z_2$ as in 
$[SU(3)]^3$ for $\eta$.  To preserve the $R_A$ parity for the gauge 
bosons, we may again define $\phi^R_{i1},\phi^R_{i2}$ to be even, and 
$\phi^R_{i3}$ to be odd.  Now $\langle \phi^R_{31} \rangle$ breaks $SU(3)_l$ 
to $SU(2)_l$,  but it also breaks $SU(2)_R$ without breaking $SU(2)_L$.  
It allows thus the separation of the $SU(2)_R$ scale without breaking 
the dark parity $R_A$.  

In the second variation (B), the electric charge is again the same as in (A) 
and the dark charge is the same as in (B) of $[SU(3)]^3$, i.e. Eq.~(11). 
Using the same changes in the pattern of symmetry breaking as discussed 
before, a model with dark $Z_2$ symmetry is again achieved.  Here 
$\langle \phi^R_{33} \rangle$ breaks $SU(3)_l \times SU(3)_R$ to 
$SU(2)_l \times SU(2)_R \times U(1)_{Y_l+Y_R}$ 
and separates the $SU(2)_l$ scale from the breaking of $SU(2)_R$ by 
$\langle \phi_{31} \rangle$.  This is the analog of the alternative 
left-right model in the $[SU(3)]^3$ case.  Applying 
$\langle \phi^L_{33} \rangle$ as well, the residual 
$U(1)$ symmetry is now $Y_L+Y_R+Y_l$, exactly as needed for the electric 
charge of Eq.~(28).  In the third variation (C), the electric charge is
\begin{equation}
Q = I_{3L} - {Y_L \over 2} + Y_R - {Y_l \over 2},
\end{equation}
and the dark charge is the same as $D_C$ of Eq.~(14).  It also results in 
a model with dark $Z_2$ symmetry.  However, as with its $[SU(3)]^3$ 
analog, it is not possible to preserve $SU(2)_L \times SU(2)_R$ as a 
low-energy subgroup.  Note that $\sin^2 \theta_W = 1/3$ at 
the unification scale for $[SU(3)]^4$ which is of order $10^{11}$ GeV 
for a nonsupersymmetric model~\cite{bmw04,kmppz17}.

\noindent \underline{\it Concluding Remarks}~:~
The existence of a dark sector is easily implemented by adding a new 
symmetry and new particles to the standard model.  There are indeed 
numerous such proposals.  As a guiding principle, supersymmetry is a 
well-known and perhaps the only example, where superpartners of all SM 
particles belong to the dark sector.  In this paper, we suggest another, 
i.e. that such a dark symmetry may have a gauge origin buried inside a 
complete extended theoretical framework for the understanding of quarks 
and leptons.  The inevitable consequence of this hypothesis is to divide 
all fermions, scalars, as well as {\it vector gauge bosons} into two 
categories.  One includes all known particles of the SM and some new ones; 
the other is the dark sector.  They are however intrinsically linked to 
each other as essential components of the unifying framework.

We consider as first examples $[SU(3)]^3$ and $[SU(3)]^4$, and identified 
the exact nature of this dark symmetry in three variations of the above 
two unified symmetries.  We have shown how 
this dark gauge symmetry is broken to the discrete $Z_2$ dark parity which 
stabilizes dark matter.  Whereas all these models contain dark matter, 
only variation (B) in either $[SU(3)]^3$ or $[SU(3)]^4$ allows it to be 
such that it exists at or near the electroweak scale.  They may serve as 
the prototypes for a deeper understanding of the origin of dark matter 
as a built-in symmetry of a theoretically motivated extension of the 
Standard Model.  Our study points to the unity of matter with dark matter, 
the origin of which is not {\it ad hoc}.  Other possible candidates are 
$SU(6)$~\cite{b12,m13} and $SU(7)$~\cite{m13}.  Future more detailed 
explorations are called for.

\noindent \underline{\it Acknowledgement}~:~
This work was supported in part by the U.~S.~Department of Energy Grant 
No. DE-SC0008541.

\bibliographystyle{unsrt}

\end{document}